# Ultimate limits to inertial mass sensing based upon nanoelectromechanical systems

K.L. Ekinci,[a)]
*Aerospace & Mechanical Engineering Department*
*Boston University, Boston, MA 02215*

Y.T. Yang, and M.L. Roukes[b)]
*Departments of Physics, Applied Physics, and Bioengineering,*
*California Institute of Technology 114-36, Pasadena, CA 91125*

Nanomechanical resonators can now be realized that achieve fundamental resonance frequencies exceeding 1 GHz, with quality factors ($Q$) in the range $10^3 \le Q \le 10^5$. The minuscule active masses of these devices, in conjunction with their high $Q$'s, translate into unprecedented inertial mass sensitivities. This makes them natural candidates for a variety of mass sensing applications. Here we evaluate the ultimate mass sensitivity limits for nanomechanical resonators operating *in vacuo* that are imposed by a number of fundamental physical noise processes. Our analyses indicate that nanomechanical resonators offer immense potential for mass sensing — ultimately with resolution at the level of *individual* molecules.

[a)] *Electronic mail: ekinci@bu.edu*
[b)] *Electronic mail: roukes@caltech.edu*

# I. INTRODUCTION

Following the trend in semiconductor electronics, mechanical devices are rapidly being miniaturized into the submicron domain.[1] Nanomechanical structures provide extremely high resonance frequencies, minuscule active masses and very small force constants. An additional and important attribute is the high quality ($Q$) factors of their resonant modes. These are currently in the range, $Q \sim 10^3$ to $10^5$, which is significantly higher than those of electrical resonant circuits. This powerful combination of attributes translates into opportunities for unprecedented mass sensitivity at high operational (resonance) frequencies — thus opening a novel realm of sensing applications.

Resonant mass sensors are already employed in many diverse fields of science and technology. These devices operate by providing a frequency shift that is directly proportional to the inertial mass of the analyte molecules accreted upon them. Among the most sensitive realizations are those based on the acoustic vibratory modes of crystals,[2,3,4] thin films[5] and micron-sized cantilevers.[6,7,8] In recent experiments,[9] we have demonstrated the unparalleled mass sensitivity of nanoelectromechanical systems (NEMS) operating *in vacuo*, using devices that take advantage of the aforementioned attributes. Given the recent realization of nanomechanical devices operating at microwave frequencies,[10] prospects for weighing *individual*, electrically neutral molecules with single-Dalton[11] sensitivity are now feasible.[9]

Two properties are central in establishing the sensitivity of resonant mass sensors: (a) the effective vibratory mass of the resonator and (b) the short- and long-term resonance frequency stability of the device. The effective vibratory mass is determined by the geometry and configuration of the resonant structure, and by the properties of the materials composing it. The frequency stability of the resonator is, in turn, governed by two classes of mechanisms — *extrinsic* processes that originate from the transducer and readout circuitry[12,13] and *intrinsic* processes fundamental to the nanomechanical resonator itself.[14,15] The frequency stability in macro- and micro-mechanical resonators have, in most cases to date, been limited by extrinsic elements. In the domain of NEMS, however, — given the enhanced sensitivity that is attainable as devices become smaller[10] and ultrasensitive displacement transduction techniques emerge[16,17] — fundamental fluctuation processes are increasingly likely to determine the outcome. These considerations motivate the present work. Below we shall explore how fundamental fluctuation processes impose ultimate limits upon the sensitivity of NEMS-based inertial mass sensors.

Even though our focus in this paper is upon establishing the fundamental limits to mass sensitivity of NEMS, the results we obtain are more general. As discussed in more detail below, a crucial aspect of investigating the ultimate mass sensitivity in NEMS is, in fact, determining the fundamental limits to frequency-shift detection. In this respect, our work is complementary to a recent paper by Cleland and Roukes,[14] in which expressions are derived for phase noise originating from a variety of physical processes. That work, however, does not address important considerations concerning specific measurement schemes that convert this phase noise into frequency fluctuations. We shall elucidate them here.

A simple and particularly illustrative structure for realizing high frequency nanomechanical resonators is the doubly-clamped beam (Fig. 1). We derive estimates for the ultimate limits to mass sensing based upon doubly-clamped Si beam resonators with a fundamental resonance frequency, $f_0 = \omega_0 / 2\pi = 1\,\text{GHz}$. We evaluate sensing prospects for two different model realizations of 1GHz fundamental-mode, doubly-clamped beam resonators; these are summarized by the parameters displayed in Table I. We use a damped, simple harmonic oscillator model to describe the flexural motion of the beam (here chosen as out-of plane, *cf.* Fig. 1) in the vicinity of the fundamental resonance. For $Q \geq 10$, this provides an accurate description of the resonant response to within 1%.[18] In this model, the mechanical response for a particular mode is approximated by that of a damped harmonic oscillator with an effective mass, $M_{eff}$, a dynamic stiffness (for point loading at the beam's center), $\kappa_{eff}$, and a quality factor, $Q$. All of these factors apply uniquely to the specific mode considered. For the *fundamental*-mode response of a simple doubly-clamped beam, the effective mass, dynamic stiffness, and the resonance frequency are given as $M_{eff} = 0.735 ltw\rho$, $\kappa_{eff} = 32Et^3 w / l^3$ and $\omega_0 = 2\pi(1.05)\sqrt{E/\rho}\,(t/l^2)$, respectively.[19] Here, $l \times t \times w$ are the beam's dimensions, $E$ is Young's modulus and $\rho$ is the mass density of the beam. We have assumed the material is isotropic; for single-crystal devices anisotropy in the elastic constants will result in a resonance frequency that depends upon specific crystallographic orientation.

In resonant sensing applications, one generally drives the resonator to a predetermined amplitude and measures amplitude[20] or frequency[13] shifts in the steady-state upon changing the resonator's physical environment. To maximize the signal-to-noise ratio (SNR) one wishes to apply the largest drive level tolerable. For the sake of concreteness, we assume that this is the maximum r.m.s. level, $\langle x_c \rangle$, still consistent with producing predominantly linear response.[21] For a doubly-clamped beam, this can be roughly approximated as $\langle x_c \rangle \approx 0.53t$, which depends only upon the beam thickness in the direction of vibration.[22] A more rigorous definition can be established for the case of frequency-stiffening nonlinearity induced by the Duffing instability for doubly-clamped beams.[23]

In the simple harmonic oscillator model, both the resonator and the added (accreted) mass, $\delta M$, are, to lowest order, approximated as point masses. Interpreting real experiments with this assumption involves consideration of additional details,[24] but these do not qualitatively alter the fundamental picture that emerges.

In Section II of this paper, we develop a detailed formalism for obtaining the mass sensitivity of a nanomechanical resonator. Mass sensitivity limits imposed by various frequency-fluctuation processes in nanomechanical resonators are discussed. Numerical estimates are obtained for the two model realizations of a 1GHz nanomechanical doubly-clamped beam resonator (Table I). In section III, we evaluate these results and present our conclusions.

## II. ANALYSIS

In general, resonant mass sensing is performed by carefully determining the resonance frequency, $\omega_0$, of the resonator and then, by looking for a frequency shift, $\delta\omega_0$, in the *steady state* due to the accreted mass. Assuming that this added mass, $\delta M$, is a small fraction of the effective vibratory resonator mass, $M_{eff}$, we can write a linearized expression,

$$\delta M \approx \frac{\partial M_{eff}}{\partial \omega_0} \delta\omega_0 = \Re^{-1} \delta\omega_0 . \tag{1}$$

This expression assumes that the modal quality factor and compliance are not appreciably affected by the accreted species. This is consistent with the aforementioned presumption that $\delta M << M_{eff}$. Hereafter, we shall refer to $\delta M$, as the *mass sensitivity* or the *minimum detectable mass* of the system. Apparently, $\delta M$ critically depends on the *minimum measurable frequency shift,* $\delta\omega_0$, and the *inverse mass responsivity,* $\Re^{-1}$.

Since the resonator's compliance, $\kappa_{eff}$, for the employed resonant mode — a function of the resonator's elastic properties and geometry — is unaffected by small mass changes, we can further determine that

$$\Re = \frac{\partial\omega_0}{\partial M_{eff}} = -\frac{\omega_0}{2M_{eff}}, \quad (2)$$

and,

$$\delta M \approx -2\frac{M_{eff}}{\omega_0}\delta\omega_0. \quad (3)$$

We note that Eq. (3) is analogous to the Sauerbrey equation,[25] but is instead here written in terms of the absolute mass, rather than the mass density, of the accreted species.

To make further progress, an important question needs to be addressed: what is the minimum measurable frequency shift, $\delta\omega_0$, that can be resolved in a (realistic) noisy system? In principle, a shift comparable to the mean square noise (the spread) in an ensemble average of a series of frequency measurements should be resolvable, *i.e.* $\delta\omega_0 \approx \frac{1}{N}\sqrt{\sum_{i=1}^{N}(\omega_i - \omega_0)^2}$ for SNR=1. An estimate for $\delta\omega_0$ can be obtained by integrating the weighted effective spectral density of the frequency fluctuations, $S_\omega(\omega)$, by the normalized transfer function of the measurement loop, $H(\omega)$ :[12]

$$\delta\omega_0 \approx [\int_0^\infty S_\omega(\omega) H(\omega) d\omega]^{1/2}. \tag{4}$$

Here, $S_\omega(\omega)$ is in units of $(rad/s^2)/(rad/s)$. We can further simplify Eq. (4), by replacing $H(\omega)$ with the square transfer function $H'(\omega)$, which has the same integrated spectral weight but is non-zero only within the passband delineated by $\omega_0 \pm \pi\Delta f$ (*i.e.* of width $\Delta f$ ; see Fig. 2). Here, $\Delta f \approx \dfrac{1}{2\pi\tau}$ and is dependent upon the measurement averaging time, $\tau$. Given this assumption, Eq. (4) takes the simpler, more familiar form

$$\delta\omega_0 \approx [\int_{\omega_0-\pi\Delta f}^{\omega_0+\pi\Delta f} S_\omega(\omega) d\omega]^{1/2}. \tag{5}$$

This, of course, is an approximation to a real system — albeit a good one. If necessary, one can resort to the more accurate expression, Eq. (4).

Equations (4) and (5) are general expressions. However, the precise functional form of $S_\omega(\omega)$ depends upon the physical noise processes that are operative, as well as the specific readout process that is employed. We shall discuss both below.

### A. Thermomechanical fluctuations

We first consider the fundamental limit imposed upon mass measurements by thermomechanical noise. These resonator fluctuations are a consequence of the fluctuation-dissipation theorem;

detailed expositions can be found elsewhere.[26] They originate from thermally-driven random motion of the mechanical device. For the one-dimensional simple harmonic oscillator representation of the NEMS modal response introduced above — characterized by $M_{eff}$ and $\kappa_{eff} = M_{eff}\omega_0^{\ 2}$ — the mean square displacement fluctuations of the center of mass, $\langle x_{th} \rangle$, satisfy $\frac{1}{2}M_{eff}\omega_0^{\ 2}\langle x_{th}^{\ 2}\rangle = \frac{1}{2}k_B T$. Here, $k_B$ is Boltzmann's constant and $T$ is the resonator temperature. The spectral density of these random displacements, $S_x(\omega)$, (with units of m$^2$/Hz) is given by

$$S_x(\omega) = \frac{1}{M_{eff}^{\ 2}} \frac{S_F(\omega)}{(\omega^2 - \omega_0^{\ 2})^2 + \omega^2\omega_0^{\ 2}/Q^2} \tag{6}$$

The thermomechanical force spectral density (with units N$^2$/Hz) has a white spectrum, $S_F(\omega) = 4M_{eff}\omega_0 k_B T/Q$.

Before proceeding, we reiterate that the manner in which Eq. (5) is evaluated will depend upon the particular measurement scheme implemented. In Figure 2, we present schematics of two common measurement circuits used for frequency tracking in mass sensing applications. In the first scheme based upon negative feedback depicted in Fig. 2(a), small phase shifts of the resonator are tracked through the use of a phase locked loop (PLL) that is driven by a constant-amplitude voltage-controlled oscillator (VCO).[27] Conversely, in the positive feedback scheme illustrated in Fig. 2(b), the resonator operates within a self-excited loop. While the final value of $\delta\omega_0$ does not depend upon the particular measurement scheme employed, the evaluation of the integral and the determination of the appropriate bandwidth does. In Fig. 2(c), we illustrate the thermomechanical noise in the resonator in relation to the measurement bandwidth of the circuit.

We now turn to the evaluation of the minimum measurable frequency shift, $\delta\omega_0$, limited by thermomechanical fluctuations of a NEMS resonator read out by a PLL circuit. In such a measurement, the resonator is driven at a constant mean square amplitude, $\langle x_c \rangle$, by the VCO. Thermal displacement fluctuations in turn generate frequency fluctuations, with an effective spectral density given by [12]

$$S_\omega(\omega) = \frac{S_\phi(\omega)}{(\partial\phi/\partial\omega)^2} \approx (\frac{\omega_0}{2Q})^2 \frac{S_x(\omega)}{\langle x_c{}^2 \rangle} \approx \frac{\omega_0^5}{Q^3} \frac{k_B T}{E_C} \frac{1}{(\omega^2 - \omega_0^2)^2 + \omega^2 \omega_0^2 / Q^2} \quad (7)$$

Here, $S_\phi(\omega)$ is the spectral density of the phase fluctuations (with units of dBc/Hz), given by $S_\phi(\omega) = \frac{S_x(\omega)}{\langle x_c{}^2 \rangle}$. We can characterize the carrier level (the VCO output delivered to the resonator) by an energy $E_c = M_{eff}\,\omega_0{}^2 \langle x_c{}^2 \rangle$, which represents the maximum drive energy. We shall, for simplicity and without loss of generality, assume this carrier to be noiseless.[28] To obtain $\delta\omega_0$, the integral in Eq. (5) must be evaluated using the expression for $S_\omega(\omega)$ given in Eq. (7) over the effective bandwidth. Performing this integration for the case where $Q>>1$ and $2\pi\,\Delta f << \omega_0 / Q$, we obtain:

$$\delta\omega_0 \approx \left[ \frac{k_B T}{E_C} \frac{\omega_0 \Delta f}{Q} \right]^{1/2} . \quad (8)$$

Both of our assumptions for evaluating the integral are reasonable. First, for a wide variety of present-day NEMS resonators $Q \geq 10^3$; second, the maximum allowable measurement bandwidth in this scheme is $\sim\omega_0/2\pi Q$ since the transient response of the resonator is characterized by a ("ring-down") time scale, $\sim Q/2\omega_0$.

The mass sensitivity is then

$$\delta M \approx 2M_{eff}\left(\frac{E_{th}}{E_c}\right)^{1/2}\left(\frac{\Delta f}{Q\omega_0}\right)^{1/2} \tag{9}$$

Here, the ratio of the maximum drive (carrier) energy, $E_c = M_{eff}\omega_0^{\,2}\left\langle x_c^{\,2}\right\rangle$ to the thermal energy, $E_{th} = k_B T$, represents the effective dynamic range intrinsic to the nanomechanical resonator itself. This is the SNR (measured in terms of power) available for resolving the coherent oscillatory response above the thermal displacement fluctuations. We can express this dynamic range, as is customary, in decibels, $DR(\text{dB}) = 10\log(E_C/k_B T)$. This yields a very simple and compelling expression

$$\delta M \approx 2M_{eff}\left(\frac{\Delta f}{Q\omega_0}\right)^{1/2} 10^{(-DR/20)}\ , \tag{10a}$$

which can be rewritten as

$$\delta M \approx \frac{1}{\Re}\left(\Delta f\,\frac{\omega_0}{Q}\right)^{1/2} 10^{(-DR/20)}\ . \tag{10b}$$

Here $\Re$ is the mass responsivity (Eq. 2), and $Q/\omega_0$ is the open-loop response ("ring-down") time of the resonator.

An identical expression to Eq. (8) can be obtained in the case of the self-excited (positive feedback) circuit of Fig. 2(b). In this case, analysis of the role of thermomechanical fluctuations in determining $\delta\omega_0$ has previously been given by Albrecht *et al.*[13] We note that the aforementioned bandwidth limitation issue, *i.e.* $\Delta f < \omega_0 / 2\pi Q$, can be circumvented by employing this self-exciting detection scheme.

In Figure 3, we plot $\delta M$ (Eq. (10)) as a function of the measurement bandwidth for the two model 1 GHz beam resonators, for two different $Q$ values. Note that mass sensitivity in the Dalton range is easily achievable, even for moderately large bandwidths. As device sizes are scaled downward while maintaining high resonance frequencies, $M_{eff}$ and $\kappa_{eff}$ shrink in direct proportion. Devices with small stiffness (high compliance) are indeed more susceptible to thermal fluctuations and consequently, the measurement dynamic range is correspondingly reduced.

**B. Temperature fluctuations**

Given its small heat capacity, a nanomechanical resonator can be subject to rather large temperature fluctuations. Its susceptibility to such fluctuations depends upon the strength of its thermal contact to the environment.[14] Since the resonator's dimensions and material parameters are both temperature dependent, temperature fluctuations will generate frequency fluctuations. Cleland and Roukes[14] have evaluated the spectral density of frequency fluctuations arising from temperature fluctuations of a NEMS resonator. They find that

$$S_\omega(\omega) = \left( -\frac{22.4 c_s^{\,2}}{\omega_0^{\,2} l^2} \alpha_T + \frac{2}{c_s} \frac{\partial c_s}{\partial T} \right)^2 \frac{\omega_0^{\,2} k_B T^2}{\pi g \left( 1 + (\omega - \omega_0)^2 \tau_T^{\,2} \right)}. \qquad (11)$$

Here, $c_s = \sqrt{E / \rho}$ is the temperature dependent speed of sound; $\alpha_T = (1/l) \partial l / \partial T$ is the linear thermal expansion coefficient; $g$ and $\tau_T$ are the thermal conductance and the thermal time constant for the nanostructure, respectively. The expression (11) was derived for a simple distributed model of thermal transport along a doubly-clamped beam of constant cross-section, neglecting material anisotropy. Upon evaluating the integral of Eq. (5), using the expression in Eq. (11) for $S_\omega(\omega)$, we obtain

$$\delta\omega_0 = \left[ \frac{1}{2\pi^2} \left( -\frac{22.4 c_s^{\,2}}{\omega_0^{\,2} l^2} \alpha_T + \frac{2}{c_s} \frac{\partial c_s}{\partial T} \right)^2 \frac{\omega_0^{\,2} k_B T^2}{g} \frac{\arctan(2\pi \Delta f \tau_T)}{\tau_T} \right]^{1/2} \quad (12)$$

and,

$$\delta M = \frac{2}{\pi^{1/2}} 2 M_{eff} \left( -\frac{22.4 c_s^{\,2}}{\omega_0^{\,2} l^2} \alpha_T + \frac{2}{c_s} \frac{\partial c_s}{\partial T} \right) \left[ \frac{k_B T^2 \arctan(2\pi \Delta f \tau_T)}{g \tau_T} \right]^{1/2} . \quad (13)$$

The values of the material dependent constants for Silicon have been calculated in Ref. 14 as $-\frac{22.4 c_s^{\,2}}{\omega_0^{\,2} l^2} \alpha_T + \frac{2}{c_s} \frac{\partial c_s}{\partial T} = 1.26 \times 10^{-4}\ /\mathrm{K}$ , $g = 7.4 \times 10^{-6}$ W/K and $\tau_T = 30\,\mathrm{ps}$. Given that $1/\tau_T$ is well above any experimental frequency shift, $\delta\omega_0$, we can approximate $\delta M$ as

$$\delta M \approx 2 M_{eff} (1.26 \times 10^{-4}\ /\mathrm{K}\ ) \left[ \frac{k_B T^2 \Delta f}{\pi g} \right]^{1/2} . \quad (14)$$

In Figure 4, we plot the above expression for $\delta M$ as a function of the measurement bandwidth for the 1 GHz doubly-clamped beams. Despite the role of thermal fluctuations in generating phase noise limitations to the mass sensitivity, single-Dalton sensing is readily achievable. Obviously, the smaller incarnations of NEMS are the most susceptible to temperature fluctuations — and this progressively becomes more significant at elevated temperatures. This can be circumvented by lowering the device temperatures and by optimizing thermal contact between a NEMS with its environment.

## C. Adsorption-desorption noise

Gas molecules in the vicinity of a resonator — each with mass, *m* — can adsorb upon the resonator's surface, *mass load* the device and, thereby, change its resonant frequency. Random, thermally-driven adsorption and desorption of molecules will therefore induce fluctuations in the resonance frequency.

This so-called adsorption-desorption noise has been discussed in detail by Yong and Vig,[29,30] and Cleland and Roukes.[14] The adsorption-desorption cycle can most conveniently be modeled by a flux-dependent adsorption rate, $r_a = \frac{2}{5}\frac{p}{\sqrt{mk_B T}}s$ and a thermally-activated desorption rate, $r_d = \nu_d \exp(-\frac{E_b}{k_B T})$. Here, *p* and *T* are the gas pressure and temperature, respectively. $E_b$ is the binding energy between the surface and the adsorbate atom. The adsorption rate, $r_a$, depends upon a phenomenological coefficient called the sticking coefficient, *s*, where $0 < s < 1$. Similarly, $r_d$ depends upon a phenomenological desorption attempt rate, $\nu_d$, where $\nu_d$ is on the order of vibrational frequencies of diatomic molecules, $\nu_d \sim 10^{13}$ Hz. Note, both $r_a$ and $r_d$ depend upon the temperature, the nature of the surface and its preparation, the adsorbing species — among other sample-specific factors. It is most convenient, therefore, to regard them as phenomenological quantities.

The spectral density of frequency fluctuations arising from adsorption-desorption processes is given by[14,29]

$$S_{\omega}(\omega) = \frac{2\pi\omega_0{}^2 N_a \sigma_{occ}{}^2 \tau_r}{\left(1 + (\omega - \omega_0)^2 \tau_r{}^2\right)} \left(\frac{m}{M_{eff}}\right)^2 . \tag{15}$$

Here, the surface is modeled as comprising $N_a$ sites for adsorption, with $\sigma_{occ}{}^2$ representing the variance in the occupation probability of a site. $\tau_r$ is the correlation time for an adsorption-desorption cycle. $\sigma_{occ}{}^2$ and $\tau_r$ can be expressed in terms of $r_a$ and $r_d$ as $\sigma_{occ}{}^2 = r_a r_d /(r_a + r_d)^2$ and $\tau_r = 1/(r_a + r_d)$, respectively.

Upon integrating $S_{\omega}(\omega)$, we obtain

$$\delta\omega_0 = \frac{1}{2\pi} \frac{m\omega_0 \sigma_{occ}}{M_{eff}} \left[N_a \arctan(2\pi\Delta f \tau_r)\right]^{1/2} . \tag{16}$$

The mass sensitivity follows as

$$\delta M \approx \frac{1}{2\pi} m\sigma_{occ} \left[N_a \arctan(2\pi\Delta f \tau_r)\right]^{1/2} . \tag{15}$$

Numerical estimates for mass sensitivity limited by adsorption-desorption noise are presented in Figure 5. The calculations were made for various background pressures of $N_2$ at 300 K. A typical sticking coefficient,[31] $s$=0.1 and a typical desorption attempt frequency,[14,29] $\nu_d = 10^{13}$ were used at $T$=300 K. We have also assumed an occupancy of approximately one adsorbate species per surface "site". The binding energy of $N_2$ on Si is $E_b$~10 kcal/mol.[31,14,29] At different pressures, the correlation time gives the cutoff frequency $1/2\pi\tau_r$ in Eq. (15).

Figure 5 clearly shows that adsorption-desorption processes will not preclude attainment of single-Dalton mass sensitivity. Adsorption-desorption noise becomes most significant in the temperature regime where the adsorption and desorption rates are comparable, hence, for a given device configuration, it can be minimized by judicious choice of operating temperature. Surface passivation to reduce the binding energy between the molecule and the surface should also be effective in this regard. Given that the adsorption-desorption noise is a surface effect, it becomes increasingly important as device sizes shrink.

## D. Momentum exchange noise

We now turn to a discussion of the consequences of momentum exchange, in a gaseous environment, between the nanomechanical resonator and the gas molecules that impinge upon it.[32] Gerlach first investigated the effect of a rarefied gas surrounding a resonant torsional mirror.[33] Subsequently, Uhlenbeck and Goudsmit[34] calculated the spectral density of the fluctuating force acting upon the mirror due to these random collisions. We reproduce here a simplified version of their discussions. In the molecular regime at *low pressure*,[32] the resonator's (representative) equation of motion takes the form[35,36]

$$M_{eff}\,\ddot{x} + \left(M_{eff}\,\frac{\omega_0}{Q_i} + \frac{pA}{\upsilon}\right)\dot{x} + M_{eff}\,\omega_0{}^2 x = F(t) \qquad (18)$$

The $(M_{eff}\,\omega_0 / Q_i)\dot{x}$ term gives rise to the intrinsic damping. The term $(pA/\upsilon)\dot{x}$ represents the drag force due to the gas molecules; $\upsilon = \sqrt{\frac{k_B T}{m}}$ is the thermal velocity of the gas molecules, $p$ the gas pressure and $A=lw$ the surface area of the beam resonator (see Fig.1). The quality factor due to the gas dissipation can be defined as $Q_{gas} = \frac{M_{eff}\,\omega_0 \upsilon}{pA}$. The so-called loaded $Q$ of the device can then be determined easily as $Q_L = (Q_U{}^{-1} + Q_{gas}{}^{-1})^{-1}$, where $Q_U$ is the intrinsic (unloaded) $Q$ of the device. Here, we focus only upon the noise due to the impinging gas molecules; we have already addressed their complement, *i.e.*, the intrinsic thermomechanical fluctuations, in section II.A. Hence, assuming that $Q_U >> Q_{gas}$ and all the fluctuations in the system result from collisions with gas molecules, the spectral density of this randomly fluctuating force, is

$$S_p(\omega) = 4m\upsilon pA = \frac{4M_{eff}\omega_0 k_B T}{Q_{gas}}. \tag{19}$$

An identical result can be deduced from Eq. (18) by using the fluctuation-dissipation theorem.

The resonator responds to this random "drive" by exhibiting displacement fluctuations with spectral density

$$S_x(\omega) \approx \frac{1}{{M_{eff}}^2} \frac{S_p(\omega)}{(\omega^2 - {\omega_0}^2)^2 + \omega^2 {\omega_0}^2 / {Q_{gas}}^2}. \tag{20}$$

Note that the form of Eq. (20) is very similar to Eq. (6) describing thermomechanical fluctuations. According to Eq. (6) these displacement fluctuations will then also appear as frequency fluctuations. After taking similar steps leading to Eq. (8), we obtain

$$\delta M \approx 2M_{eff}\left(\frac{E_{th}}{E_c}\right)^{1/2}\left(\frac{\Delta f}{Q_{gas}\omega_0}\right)^{1/2}. \tag{21}$$

Figure 6 shows $\delta M$ for the representative 1 GHz resonators at atmospheric pressures of $N_2$ at 300 K. The inset to Figure 6 shows how $Q_{gas}$ evolves as a function of the gas pressure. Momentum noise appears to be an insignificant source of noise for NEMS operating *in vacuo*, *i.e.* $p$<1 mTorr.

## III. DISCUSSION

In Section II we have evaluated the ultimate sensitivity limits to mass sensing via nanomechanical resonators that are imposed by several important noise processes. Our analysis culminates in the expression Eq. (10b),

$$\delta M \approx \frac{1}{\Re}\left(\Delta f\,\frac{\omega_0}{Q}\right)^{1/2} 10^{(-DR/20)} \tag{22}$$

and its equivalent, Eq. (10a). Eq. (22) distills and makes transparent the essential considerations for optimizing inertial mass sensors – at any size scale. There are three principal considerations. First, the mass responsivity, $\Re$, should be maximal. As seen from Eq. (2), this emphasizes the importance of devices possessing low mass, *i.e.* small volume, which operate with high resonance frequency. Second, the measurement bandwidth should employ the full range that is available.[37] Third, the dynamic range for the measurement should be maximized. At the outset, this latter consideration certainly involves careful engineering to minimize what we have termed "extrinsic" noise processes. But this is ultimately feasible only down to the point where fundamental limits are reached. In this regime it is the "intrinsic" noise processes that become predominant. In this paper we have evaluated those that are most important; Table II summarizes the corresponding functional forms of noise-limited mass sensitivity that result.

In Figs. 2 through 6, we have translated these analytical results into concrete numerical estimates for the two representative, and *realizable* configurations of 1GHz doubly-clamped beam silicon resonators. The implications of these results are manifest – the plot abscissae span only the regime from a few tenths, to a few tens of Daltons. This is the mass range for a small *individual* molecule; hence it is clear that nanomechanical mass sensors offer unprecedented sensitivity.

This raises an important question: how optimally can a nanomechanical device perform for sensing species with large mass, say in the megadalton (MDa, *i.e.* $10^6$ Da) range? Heretofore, our focus has solely been upon the "mass noise floor", and not on how much can be accreted upon a NEMS mass sensor without degrading its performance. We will not here carry out a detailed analysis of the attainable "mass dynamic range" of nanomechanical sensors, but instead offer several brief comments. First, masses of the resonators we consider are themselves of order 0.1 to 10 fg ($10^8$ to $10^{10}$ Da) – *i.e.* much greater than that of an individual 1 MDa particle. Accretion of hundreds, to hundreds of thousands of such macromolecules would only shift the natural (unloaded) resonant frequency downward by about 50%. At this surface coverage, the molecules, which are much more mechanically compliant than silicon, would negligibly perturb the dynamic stiffness of the resonator, $\kappa_{eff}$. In fact, for surface coverage below one monolayer, one would also expect the resonator's quality factor to be minimally affected by the adsorbates. Accordingly, in addition to their unprecedented mass sensitivity, NEMS mass sensors appear to offer remarkably large mass dynamic range. Our recent experiments, reported elsewhere, confirm this.[9]

In these recent mass sensitivity measurements,[9] extrinsic amplifier noise processes have imposed the dominant source of phase noise. Recently, significant progress has been made in pushing this phase noise down close to fundamental limits.[38] With such advances, it is clear that nanomechanical mass sensing with the single-Dalton sensitivity will be realizable in the near future. This will give researchers the unprecedented ability to *weigh* individual neutral molecules routinely — blurring the distinction between conventional inertial mass sensing and mass spectrometry.

## IV. ACKNOWLEDGEMENTS

YTY and MLR are grateful to DARPA MTO/MEMS and SPAWAR for supporting this work under grant N66001-01-X-6004/ 02-8914/1000000928. The authors acknowledge many fruitful conversations with A. Vandelay.

**FIGURE CAPTIONS:**

**FIGURE 1.** Doubly-clamped beam resonator with length, *l*, width, *w*, and thickness, *t*. Our illustrative analyses are based upon the fundamental-mode, out-of–plane (*z*-directed) flexural response of the beam.

**TABLE I.** Parameters for the two representative 1 GHz doubly clamped beams considered in this work.

**FIGURE 2.** Schemes for the operation of two port resonant NEMS devices. **(a)** In the phase locked loop (PLL), the principal components are: (**VCO**) voltage controlled RF oscillator; (**PS**) power splitter; two port-**NEMS**; (**M**) mixer (with RF, LO and IF ports); (**Ø**) phase shifter; (**L**) amplitude limiter; (**A**) variable gain amplifier; (**LPF**) low pass filter. **(b)** In a self excited circuit, similar components are used. **(c)** Power spectral density of thermomechanical fluctuations in a driven simple harmonic oscillator as a function of frequency, normalized to the response on resonance for *Q*=1. The coherent drive, assumed noiseless, is represented by the vertical arrow at the resonance frequency, $\omega = \omega_0$. Thermomechanical noise in the measurement bandwidth contributes to the observed frequency fluctuations. A large measurement bandwidth, $\Delta f$, in general, results in enhanced noise, but gives a better characteristic response time, $\tau = 1/(2\pi \Delta f)$. The measurement bandwidth in the phase locked loop (PLL) measurement, for instance, is determined either by the bandwidth of the low pass loop filter (LPF) that is employed, or by the ring-down time of the resonator itself.

**FIGURE 3.** Limits to mass sensitivity, $\delta M$, imposed by thermomechanical fluctuations, in units of Daltons (Da), as a function of the measurement bandwidth, $\Delta f$, for the two representative 1 GHz resonators described in the text. Here, for each device, $\delta M$ is presented for two different values of *Q*. Although the ordinate extends to $10^7$ Hz, note that the attainable open-loop measurement bandwidth, $\omega_0/(2\pi Q)$, is limited to ~$10^6$ and ~$10^4$ Hz, for $Q = 10^3$ and $10^5$, respectively. The attainable bandwidth will be further altered in a closed-loop measurement (*i.e.* with feedback).

**FIGURE 4.** Mass sensitivity limits imposed by temperature fluctuations as a function of measurement bandwidth, for the two representative 1 GHz silicon resonators, for operation at *T*=300 K. The accessible measurement bandwidth is subject to the same restrictions mentioned in connection with Fig. 3.

**FIGURE 5.** Limits to mass sensitivity imposed by adsorption-desorption processes, for the two representative 1 GHz doubly-clamped silicon beam resonators described in the text. The calculations displayed are for three different pressures of $N_2$ with *s*=0.1 and $\nu_d = 10^{13}$ (see text) — with approximately one adsorbate per surface silicon atom site. The accessible measurement bandwidth is subject to the same restrictions mentioned in connection with Fig. 3.

**FIGURE 6.** Limits to mass sensitivity set by momentum exchange noise between the resonator and gas molecules for a resonator intrinsic *Q* of $Q_U = 10^5$ at atmospheric pressure of $N_2$, *p*=760 Torr. The inset shows $Q_{gas}$ as a function of the gas pressure for both resonators. The momentum exchange noise becomes relevant only when $Q_U >> Q_{gas}$, i.e. for $p$>>1 Torr for the Si nanowire resonator and $p$>>10 Torr for the Si beam resonator. The accessible measurement bandwidth is subject to the same restrictions mentioned in connection with Fig. 3.

**Figure 1.**

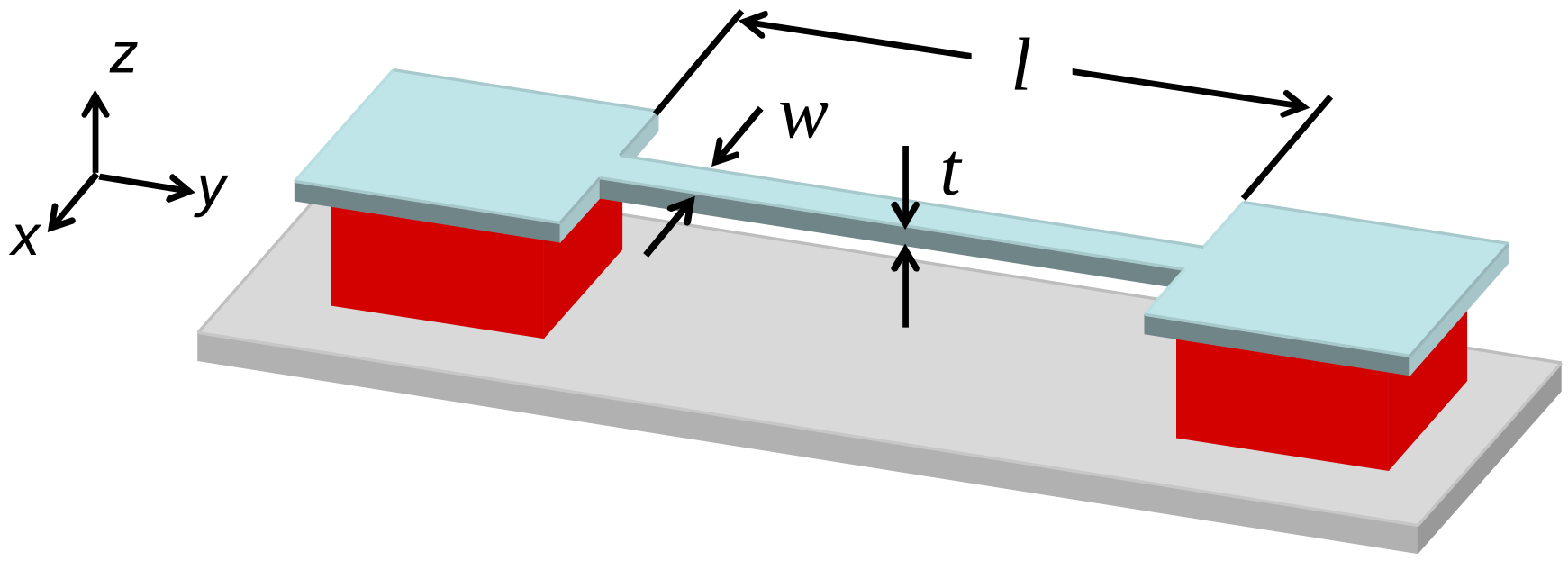

**FIGURE 1.** Doubly-clamped beam resonator with length, *l*, width, *w*, and thickness, *t*. Our illustrative analyses are based upon the fundamental-mode, out-of–plane (*z*-directed) flexural response of the beam.

**Table I.**

**TABLE I.** Parameters for the two representative 1 GHz doubly clamped beams considered in this work.

| $w \times t \times \ell$ (nm) | $M_{eff} = 0.735 wlt\rho$ (g) | $\kappa_{eff}$ (N/m) | $\langle x_c \rangle$ (nm) | $E_c = M_{eff}\omega_0^2\langle x_c^2 \rangle$ (J) | *DR* at 300 K $DR = 10\log(E_c/k_B T)$ (dB) |
|---|---|---|---|---|---|
| $50 \times 80 \times 780$ Si Beam | $5.30\mathrm{x}10^{-15}$ | ~290 | 42 | $3.7\mathrm{x}10^{-13}$ | ~80 |
| $15 \times 15 \times 340$ Si nanowire | $1.30\mathrm{x}10^{-16}$ | ~6.73 | 8 | $3.5\mathrm{x}10^{-16}$ | ~50 |

**Figure 2.**

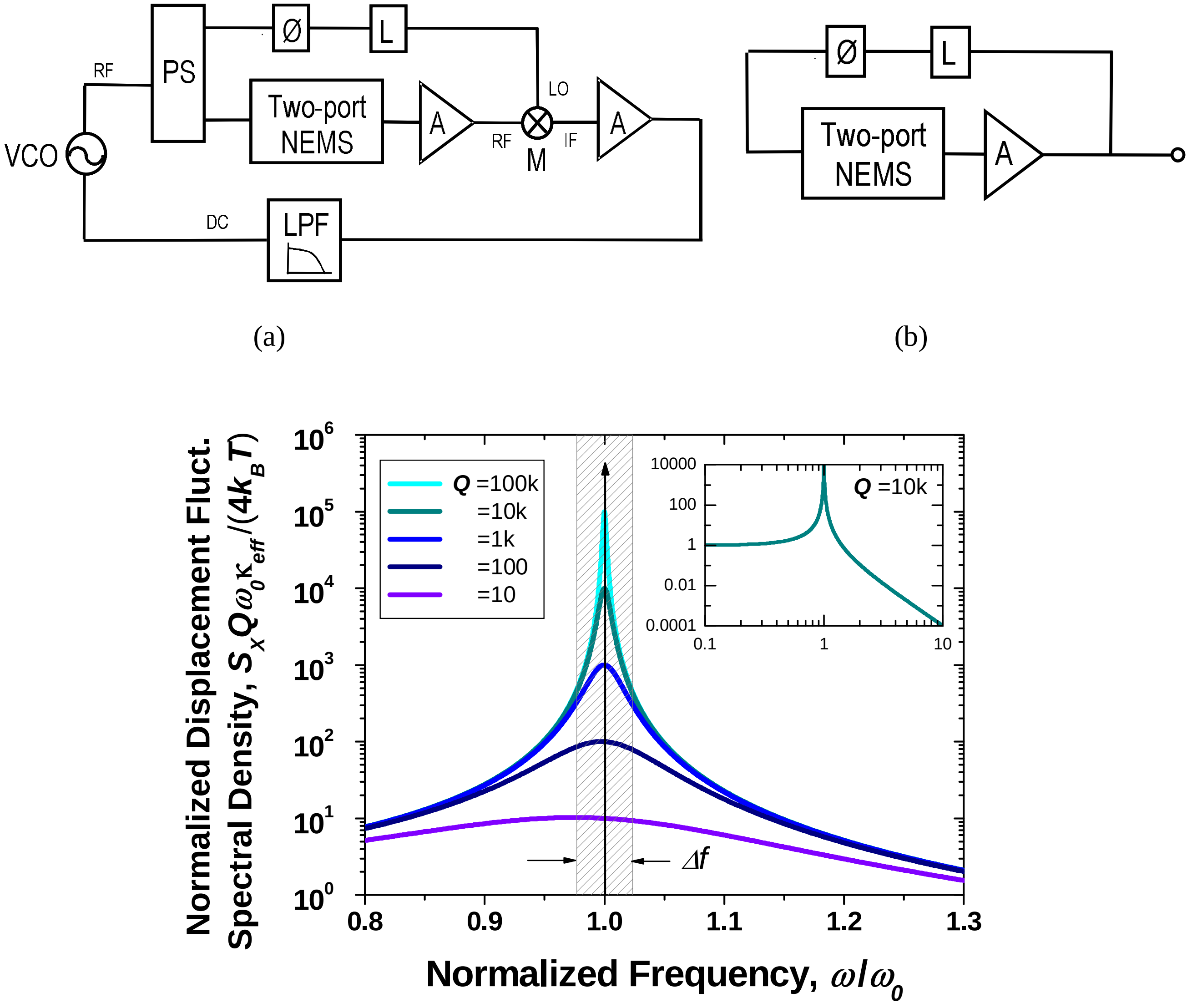

(c)

**FIGURE 2.** Schemes for the operation of two port resonant NEMS devices. **(a)** In the phase locked loop (PLL), the principal components are: (**VCO**) voltage controlled RF oscillator; (**PS**) power splitter; two port-**NEMS**; (**M**) mixer (with RF, LO and IF ports); (**Ø**) phase shifter; (**L**) amplitude limiter; (**A**) variable gain amplifier; (**LPF**) low pass filter. **(b)** In a self excited circuit, similar components are used. **(c)** Power spectral density of thermomechanical fluctuations in a driven simple harmonic oscillator as a function of frequency, normalized to the response on resonance for $Q$=1. The coherent drive, assumed noiseless, is represented by the vertical arrow at the resonance frequency, $\omega = \omega_0$. Thermomechanical noise in the measurement bandwidth contributes to the observed frequency fluctuations. A large measurement bandwidth, $\Delta f$, in general, results in enhanced noise, but gives a better characteristic response time, $\tau = 1/(2\pi\Delta f)$. The measurement bandwidth in the phase locked loop (PLL) measurement, for instance, is determined either by the bandwidth of the low pass loop filter (LPF) that is employed, or by the ring-down time of the resonator itself.

**Figure 3.**

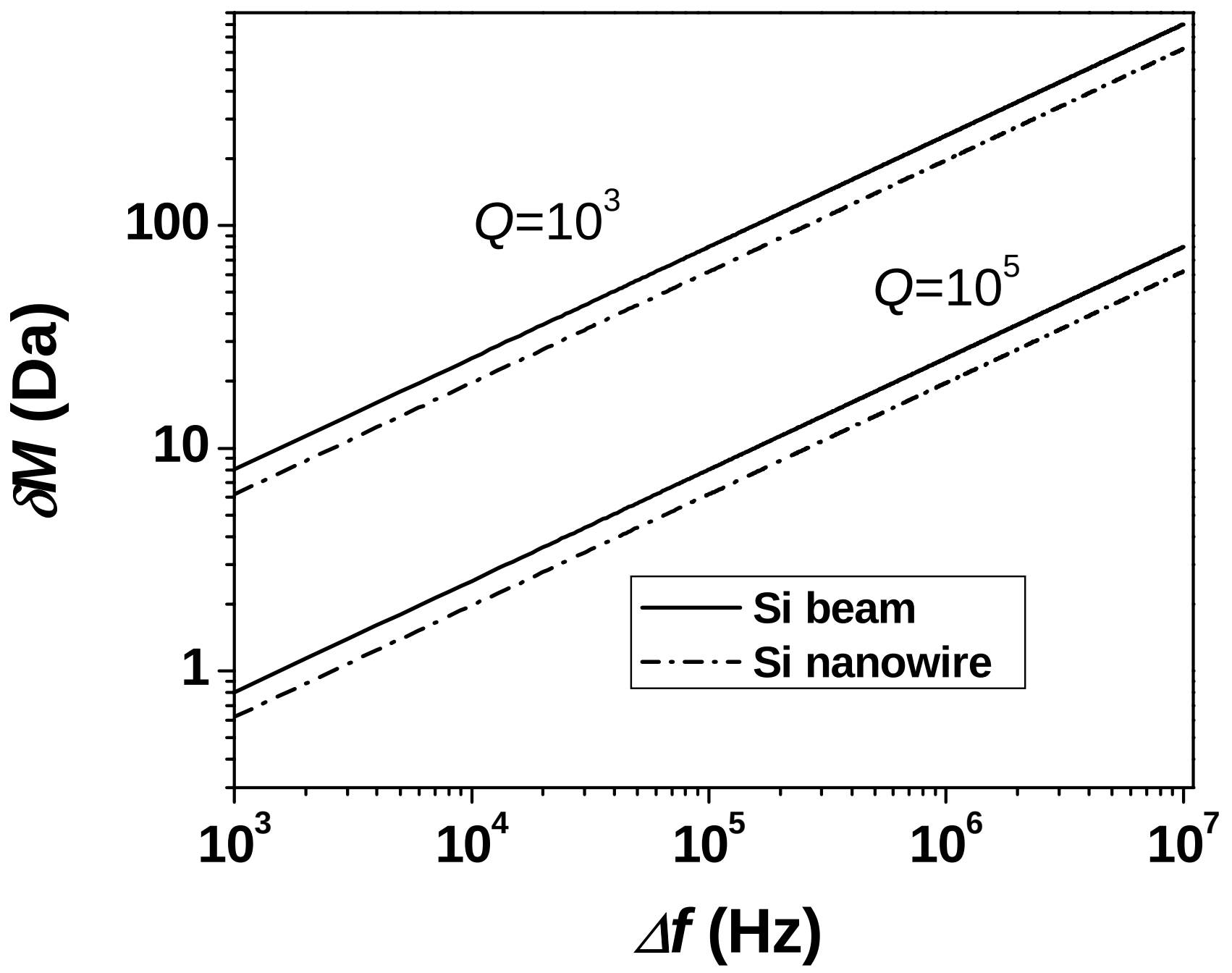

**FIGURE 3.** Limits to mass sensitivity, $\delta M$, imposed by thermomechanical fluctuations, in units of Daltons (Da), as a function of the measurement bandwidth, $\Delta f$, for the two representative 1 GHz resonators described in the text. Here, for each device, $\delta M$ is presented for two different values of $Q$. Although the ordinate extends to $10^7$ Hz, note that the attainable open-loop measurement bandwidth, $\omega_0/(2\pi Q)$, is limited to ~$10^6$ and ~$10^4$ Hz, for $Q$ = $10^3$ and $10^5$, respectively. The attainable bandwidth will be further altered in a closed-loop measurement (*i.e.* with feedback).

**Figure 4.**

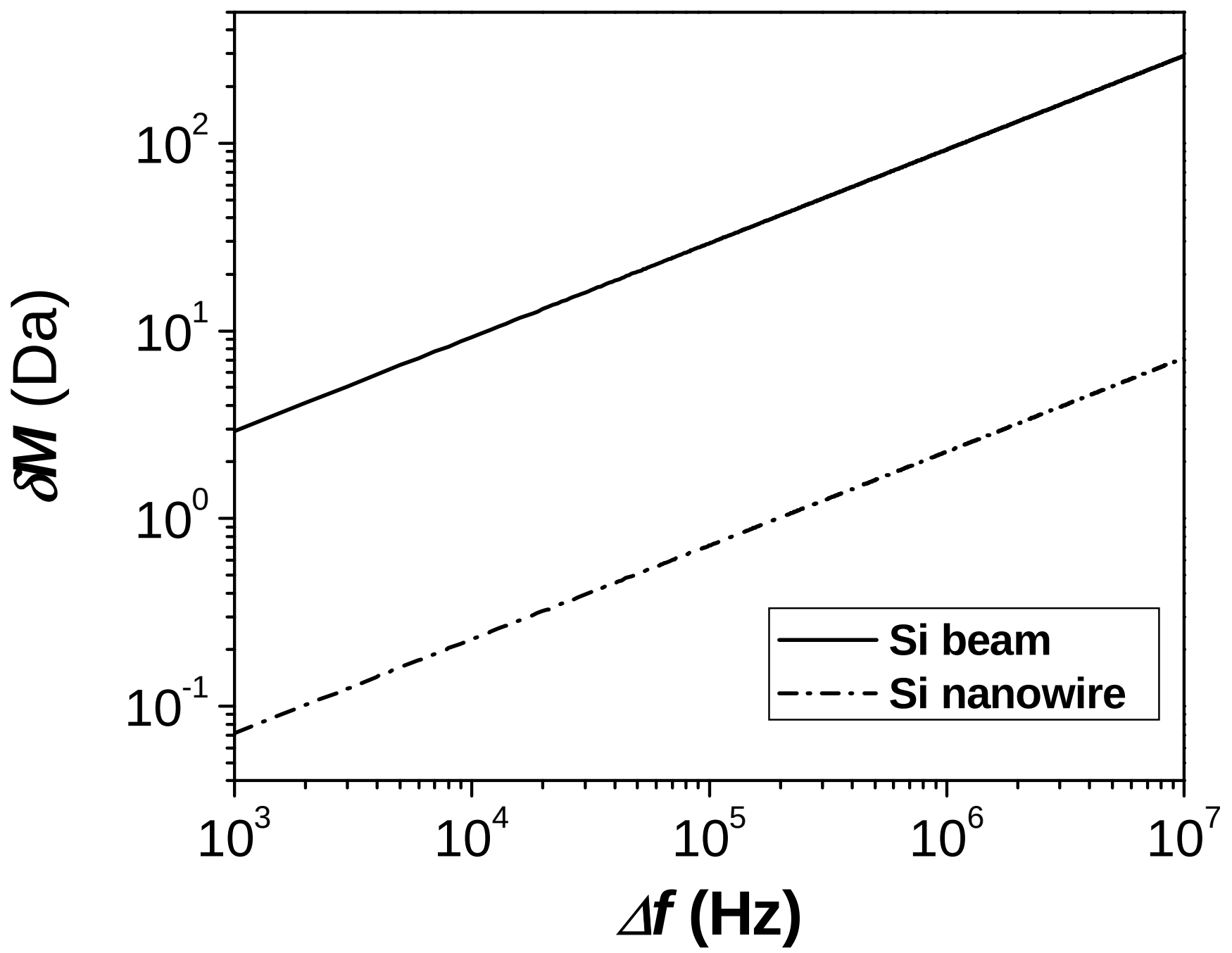

**FIGURE 4.** Mass sensitivity limits imposed by temperature fluctuations as a function of measurement bandwidth, for the two representative 1 GHz silicon resonators, for operation at *T*=300 K. The accessible measurement bandwidth is subject to the same restrictions mentioned in connection with Fig. 3.

**Figure 5.**

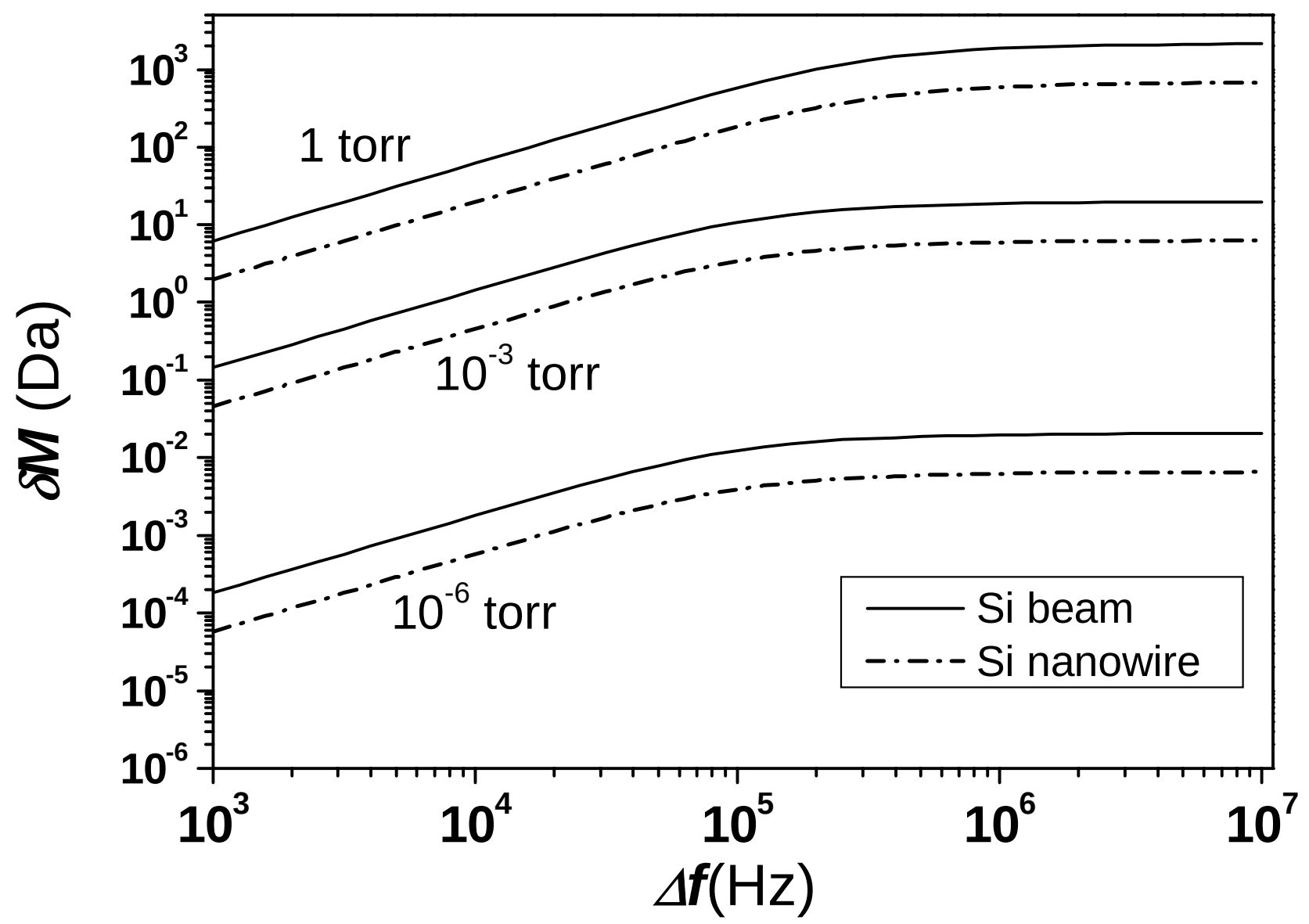

**FIGURE 5.** Limits to mass sensitivity imposed by adsorption-desorption processes, for the two representative 1 GHz doubly-clamped silicon beam resonators described in the text. The calculations displayed are for three different pressures of $N_2$ with $s$=0.1 and $\nu_d = 10^{13}$ (see text) — with approximately one adsorbate per surface silicon atom site. The accessible measurement bandwidth is subject to the same restrictions mentioned in connection with Fig. 3.

**Figure 6.**

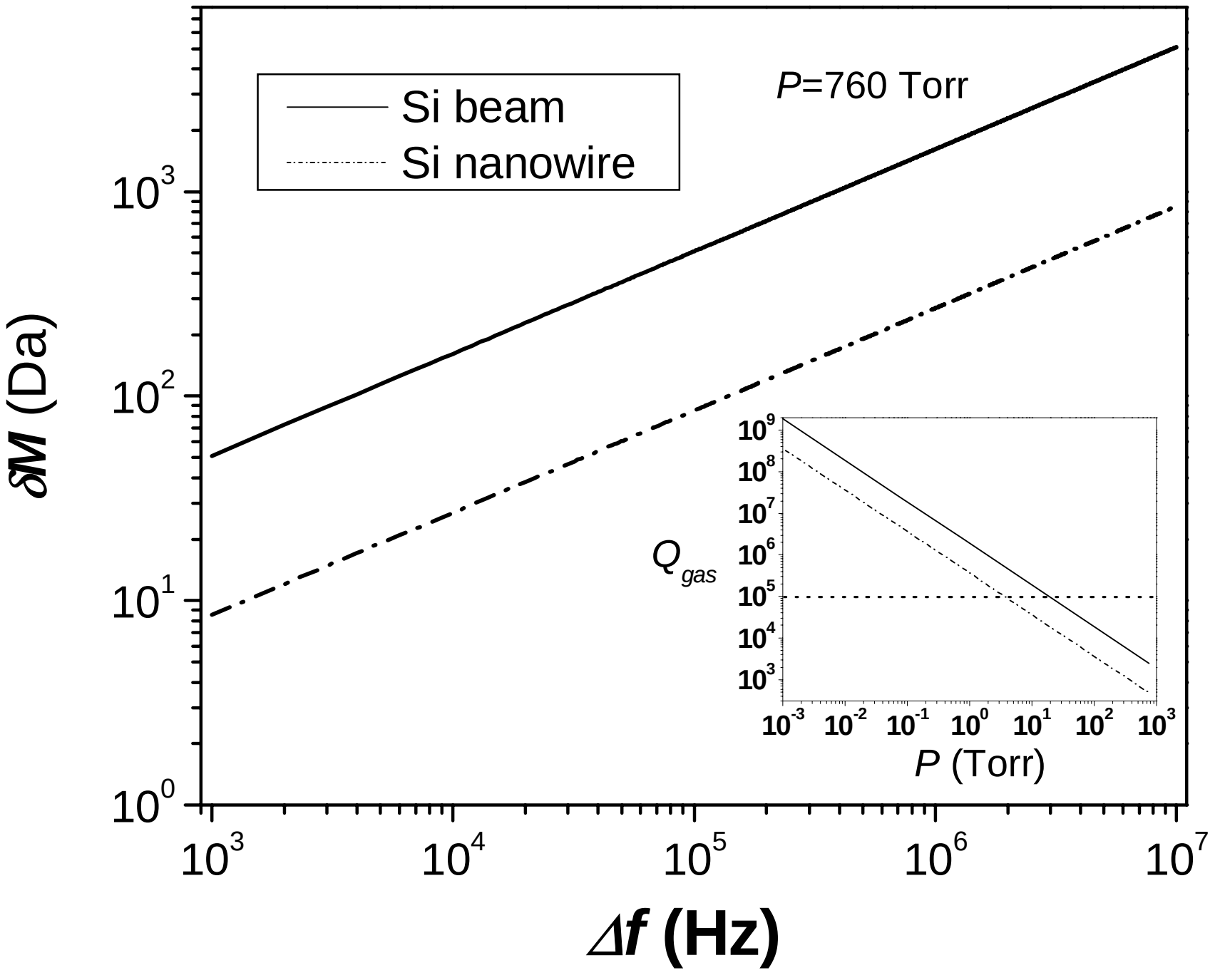

**FIGURE 6.** Limits to mass sensitivity set by momentum exchange noise between the resonator and gas molecules for a resonator intrinsic $Q$ of $Q_U = 10^5$ at atmospheric pressure of $N_2$ ($p$=760 Torr). The inset shows $Q_{gas}$ as a function of the gas pressure for both resonators ($Q_U$=$10^5$ is shown by the horizontal dotted line). The momentum exchange noise becomes relevant only when $Q_{gas} \ll Q_U$, i.e. for $p \gg 1$ Torr for the Si nanowire resonator and $p \gg 10$ Torr for the Si beam resonator. The accessible measurement bandwidth is subject to the same restrictions mentioned in connection with Fig. 3.

**Table II.**

**TABLE II.** Expressions for mass sensitivity for different physical noise mechanisms.

| Type of noise | $S_\omega(\omega)$ | $\delta M$ |
|---|---|---|
| Thermomechanical | $\frac{\omega_0^5}{Q^3}\frac{k_B T}{E_C}\frac{1}{(\omega^2-\omega_0^2)^2+\omega^2\omega_0^2/Q^2}$ (PLL)<br>$\frac{k_B T\omega_0}{E_c Q}$ (self-excited loop) | $2M_{eff}\left(\frac{k_B T}{E_c}\right)^{1/2}\left(\frac{\Delta f}{Q\omega_0}\right)^{1/2}$ |
| Temperature fluctuations | $\left(-\frac{22.4c_s{}^2}{\omega_0{}^2 l^2}\alpha_T+\frac{2}{c_s}\frac{\partial c_s}{\partial T}\right)^2\frac{\omega_0{}^2 k_B T^2}{\pi g\left(1+(\omega-\omega_0)^2\tau_T{}^2\right)}$ | $\frac{4M_{eff}}{\pi^{1/2}}\left(-\frac{22.4c_s{}^2}{\omega_0{}^2 l^2}\alpha_T+\frac{2}{c_s}\frac{\partial c_s}{\partial T}\right)\left[\frac{k_B T^2\arctan(2\pi\Delta f\tau_T)}{g\tau_T}\right]^{1/2}$ |
| Adsorption-desorption | $\frac{2\pi\omega_0{}^2 N_a\sigma_{occ}{}^2\tau_r}{\left(1+(\omega-\omega_0)^2\tau_r{}^2\right)}\left(\frac{m}{M_{eff}}\right)^2$ | $\frac{1}{2\pi}m\sigma_{occ}\left[N_a\arctan(2\pi\Delta f\tau_r)\right]^{1/2}$ |
| Momentum exchange | $\frac{\omega_0^5}{Q_{gas}{}^3}\frac{k_B T}{E_C}\frac{1}{(\omega^2-\omega_0^2)^2+\omega^2\omega_0^2/Q_{gas}{}^2}$ (PLL)<br>$\frac{k_B T\omega_0}{E_c Q_{gas}}$ (self-excited loop) | $2M_{eff}\left(\frac{k_B T}{E_c}\right)^{1/2}\left(\frac{\Delta f}{Q_{gas}\omega_0}\right)^{1/2}$ |

---

---

[32] We will only consider the *low pressure limit*. The expressions in the text will not be valid for high pressures, where the dissipation is due to the viscous forces of the fluid. Yet, the crossover to the viscous regime for small, high frequency resonators takes place at higher pressures. This crossover pressure can be determined by comparing the sound wavelength in the medium to the mean free path of the gas molecules. The mean free path, $\lambda_{mfp}$, in a gaseous environment with molecules modeled as hard spheres is given by $\lambda_{mfp} = \frac{k_B T}{\sqrt{2}\pi d_0^{\,2} p}$, where $d_0$ is the sphere diameter, $T$ the temperature and $p$ the pressure of the gas; $k_B$ is Boltzmann's constant. Wavelength of the sound waves around the operation frequency of the resonator can be estimated from a linear dispersion. One can then determine that for a nanomechanical resonator at 1GHz, this crossover pressure is larger than atmospheric pressure. For more details, see, for example, V.B. Braginsky, V.P. Mitrofanov and V.I. Panov, *Systems with Small Dissipation* (The University of Chicago Press, Chicago, 1985).

[33] W. Gerlach, Naturwiss. **15**, 15 (1927).

[34] G.E Uhlenbeck, and S. Goudsmit , Phys. Rev. **34**, 145 -151(1929).

[35] K. Kokubun, M. Hirata, M. Ono, H. Murakami, and Y. Toda, J. Vac. Sci. Technol. **A 5**, 2450 (1987).

[36] F.R. Blom, S. Bouwstra, M. Elwenspoek, J.H.J. Fluitman, J. Vac. Sci. Technol. **B 10**, 19 (1992).

[37] In the presence of feedback, this consideration is modified to involve effective (closed-loop) $Q$, rather than the natural (open-loop) $Q$, since the former will then determine the bandwidth.

[38] Y.T. Yang, C. Callegari, X.L. Feng, and M.L. Roukes, *to be published.*